\documentclass[twocolumn,showpacs,preprintnumbers,amsmath,amssymb]{revtex4}

\usepackage{graphicx}

\begin{document}


\title{Thermal equilibrium and efficient evaporation of an ultracold atom-molecule mixture}

\author{Cheng Chin and Rudolf Grimm}

\affiliation{Institut f\"{u}r Experimentalphysik, Universit\"{a}t
Innsbruck, Technikerstr. 25, 6020 Innsbruck, Austria}

\date{\today}

\begin{abstract}
We derive the equilibrium conditions for a thermal atom-molecule
mixture near a Feshbach resonance. Under the assumption of low
collisional loss, thermodynamical properties are calculated and
compared to the measurements of a recent experiment on fermionic
lithium experiment \cite{joc03}. We discuss and evaluate possible
collision mechanisms which can lead to atom-molecule conversion.
Finally, we propose a novel evaporative cooling scheme to
efficiently cool the molecules toward Bose-Einstein condensation.
\end{abstract}

\pacs{34.50.-s, 05.30.Fk, 39.25.+k}


\maketitle\narrowtext\section{Introduction}

Recent experiments witness a fast and remarkable progress in
creating molecules from atomic Bose-Einstein condensates (BEC)
\cite{don02, her03, dur03}, degenerate Fermi gases \cite{reg03,
cub03, str03} or ultracold thermal gases \cite{chi03, joc03}.
These results have initiated new pursuits toward molecular BEC,
Cooper-paired Fermi gas \cite{hol01, oha02} as well as matter-wave
interferometry based on ultracold molecules.

In all these experiments, the creation of molecules is based on
magnetically tunable Feshbach resonances \cite{fes62, ino98},
which allow interacting atom pairs to couple to molecules in a
single internal quantum state. Extensive studies on atomic
Feshbach resonances not only lead to a thorough understanding of
the cold atom collision properties, but also suggest intriguing
possibility to convert atomic BEC or degenerate Fermi gas to a
molecular BEC with high efficiency \cite{tim99, van99, mie00,
tim01}.

Converting atoms into molecules via a Fesh\-bach resonance is
accompanied by other effects: three-body collisions \cite{fed96,
esr01, pet03}, atom-molecule interactions \cite{cub03},
molecule-molecule interactions \cite{joc03}, as well as many-body
effects in quantum degenerate gases \cite{don02}. These processes
can lead to a low conversion efficiency or a short molecular
lifetime, both limit the ability to observe and manipulate the
molecules. As the detailed mechanisms of the conversion process
require future investigations, it remains an experimental task to
identify the best system and strategy to reach molecular
condensation.

In most recent experiments, molecules are created by ramping the
magnetic field through a Feshbach resonance. High conversion
efficiencies of $50\%$ to $80\%$ are found in degenerate Fermi
gases \cite{reg03, cub03, str03}, which agree with the
calculations \cite{kok03, hu03}. Surprisingly, a recent experiment
showed that a similar efficiency can be achieved in a thermal gas
of fermionic $^6$Li atoms at fixed magnetic field \cite{joc03}.
Furthermore, the molecule sample can be purified, trapped and
remains stable with a lifetime of up to 10s. The results of Ref.
\cite{joc03} prompt us to investigate the thermodynamics of a
non-degenerate atom-molecule mixture and the possibility to cool
the molecules to a molecular BEC.

In this paper, we introduce a thermodynamical model to calculate
the atom-molecule conversion efficiency in a thermal gas (Sec.
II). Using the model, we derive the atom-molecule (molecule-atom)
conversion efficiency and compare the results with the
experimental data (Sec. III and IV). To further cool the
molecules, we propose a novel evaporative cooling scheme in an
atom-molecule mixture (Sec. V). Finally, we identify the relevant
conversion mechanisms and calculate the time scale of the proposed
cooling process (Sec. VI).





\section{Model}
Based on the fermionic lithium system, we consider a classical
thermal sample with $N$ and $N'$ atoms in two internal states and
$M$ molecules. The molecules individually consist of two
non-identical atoms and are in a single internal state with
constant binding energy $E$. We assume the system has negligible
collision loss and energy exchange with the environment.

The conservation of particle number and energy imposes two
constraints in the thermalization process: the former is simply
$N+M=const.$ and $N'+M=const.$, while the latter is given by
$Ne_n+N'e_{n'}+M(e_m-E)=const.$, where $e_n$ $(e_{n'})$ and $e_m$
are the mean external energy per atom and molecule. We assume the
gas is dilute with negligible interaction energy among particles.

The thermal equilibrium condition is obtained assuming that the
atom-molecule mixture is in a quantum-mechanical canonical
ensemble. This assumption is generally valid in the limit of large
particle number \cite{kerson}. Given the temperature $T$ and
single-particle partition function $Z_n$ $(Z_{n'})$ and
$Z_me^{-E/kT}$ for atoms and molecules, we derive the equilibrium
condition by minimizing the free energy $F=-kT$ ln$Z$, subjected
to only the particle number conservation constraint, where
$Z=Z_n^NZ_{n'}^{N'}Z_m^Me^{-ME/kT}(N!N'!M!)^{-1}$ is the partition
function of the system, and $k$ Boltzmann's constant. The result
constitutes the key equation we investigate in this paper,

\begin{eqnarray}
\phi_M=\phi_N\phi_{N'}e^{E/kT}. \label{eq1}
\end{eqnarray}

\noindent Here $\phi_N=N/Z_n$ $(\phi_M=M/Z_m)$ is the final atomic
(molecular) phase-space density in the motional ground state.

This equation is valid for systems with atoms in two internal
states and molecules in one state, as well as those with two
atomic species, say, Rb atoms, Cs atoms with RbCs molecules. The
only assumptions are the low collision loss and interaction
energy. For single component systems with $N$ atoms and $M$
molecules, Eq.~(\ref{eq1}) is rewritten as
$\phi_M=\phi_N^2e^{E/kT}$ \cite{chi03}.

\begin{figure}
\includegraphics[width=2.75in]{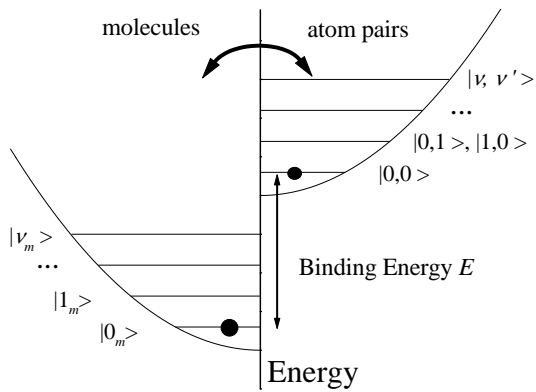}
\caption{Thermal equilibrium in an atom-molecule mixture.
Molecular states $|\nu_m\rangle$ are introduced to the two-body
states of non-identical atom pairs $|\nu,\nu'\rangle$, where
$\nu$, $\nu'$ $(\nu_m)$ are the motional quantum numbers of a
single atom (molecule). Population in $|0,0\rangle$ is the product
of that in single-particle state $|\nu=0\rangle$ and
$|\nu'=0\rangle$, namely, $\phi_N\phi_{N'}$. Molecular phase-space
density is enhanced by the Boltzmann factor $e^{E/kT}$ when
$E>0$.} \label{fig0}
\end{figure}

The thermalization condition given in Eq.~(\ref{eq1}) can be
understood in a simple picture. As the atomic phase-space density
$\phi_N$ and $\phi_{N'}$ are the ground state populations of the
two atomic components, $\phi_N\phi_{N'}$ is the population of the
non-identical atom pairs in the lowest two-atom state.
Eq.~(\ref{eq1}) states that the motional ground state of the
molecule is regarded as one of the two-atom states. An additional
Boltzmann factor $e^{E/kT}$ accounts for the binding energy. This
is illustrated in Fig.~\ref{fig0}. Notice that this result is
independent of the details and complexity of the molecule
formation and dissociation mechanisms.

Interesting consequences can be directly observed from
Eq.~(\ref{eq1}). At constant temperature, molecular density is
proportional to atomic density squared, even in the case when the
molecules are formed, say, by a three-body process, whose rate
depends on density cubed. This is because the reverse molecule
dissociation rate also depends on atomic density. The balance of
both the formation and dissociation processes in thermal
equilibrium gives the density squared dependence. We will come
back to the dissociation process in Sec. VI.

In the following sections, we calculate the atom (molecule) number
$N$ $(M)$ and temperature $T$ in thermal equilibrium based on
Eq.~(\ref{eq1}) and the conservation laws: we first determine the
constants of the conservation laws from the initial conditions,
then express partition function $Z_n$, $Z_ {n'}$ and $Z_m$ in
terms of trapping potential and temperature $T$; mean energy in
terms of partition function and temperature,
$e_i=-Z_i^{-1}\partial_{\beta} Z_i$, where $\beta=(kT)^{-1}$. In
cases where no analytic solutions are available, we solve the
equations numerically.

A special case is calculated here when the molecules and two
atomic components are identically trapped $Z_n=Z_{n'}=Z_m$ with
initial atom numbers $N_0$, $N'_0$ and molecule number $M_0=0$.
The conversion efficiency $f=2M/(N_0+N'_0)$ can be derived from
Eq.~(\ref{eq1}) and particle conservation law as

\begin{eqnarray}
f=\left(1+\frac{\phi_N+\phi_{N'}}{2\phi_N\phi_{N'}}
e^{-E/kT}\right)^{-1}. \label{eq2}
\end{eqnarray}

When the atomic gas is tuned right on resonance $E=0$ with final
phase-space density $\phi_N=\phi_{N'}=1$, an conversion efficiency
of $50\%$ is obtained. Notably, the unity phase-space density
assumption is on the boarder of the applicability of our model.

On the other hand, given an atom sample with low initial
phase-space densities $\phi_{N_0}=\phi_{N'_0}\ll 1$ and zero
binding energy, the conversion efficiency is simply
$f=\phi_N=\phi_{N'} \sim \phi_{N_0}$. This result indicates that
creating molecules from a cloud of thermal atoms by tuning the
field right on Feshbach resonance is inefficient. In the following
sections, we show that an appreciable gain in conversion fraction
can be obtained when we tune the molecular state below continuum.

\section{Converting atoms to molecules}

The atom-molecule conversion can be qualitatively understood from
Eq.~(\ref{eq1}). When the molecular state is far below (far above)
the atomic continuum, all particles should accumulate in the lower
molecular (atomic) state and the atom (molecule) number is
exponentially suppressed. This naive picture, however, is
incorrect when one tries to convert cold atoms into molecules. At
large binding energy, the internal energy released during the
molecule formation process heats up the sample significantly and
reduces the atomic phase-space density. In the following
calculation, we show that the final temperature goes up
approximately linearly with the binding energy and the molecule
fraction is therefore limited. In the limit of infinite binding
energy, counter-intuitively, no molecules are formed.

To show the conversion in the vicinity of a Feshbach resonance, we
consider both atoms and molecules are harmonically trapped with
identical single-particle partition function,
$Z_n=Z_{n'}=Z_m=\prod_i(1-e^{-\hbar\omega_i/kT})^{-1}$, where
$\omega_i$ is the trap vibration frequency in the $i$th direction.
The external energy of the particles is $e_n=e_n'=e_m=3kT$. The
assumption that both species have the same trap vibration
frequencies is generally valid for atoms and long-range atomic
dimers in a deep far-detuned dipole trap where the trap depth and
mass for a molecule are both twice as large as those for an atom.

\begin{figure}
\includegraphics[width=3in]{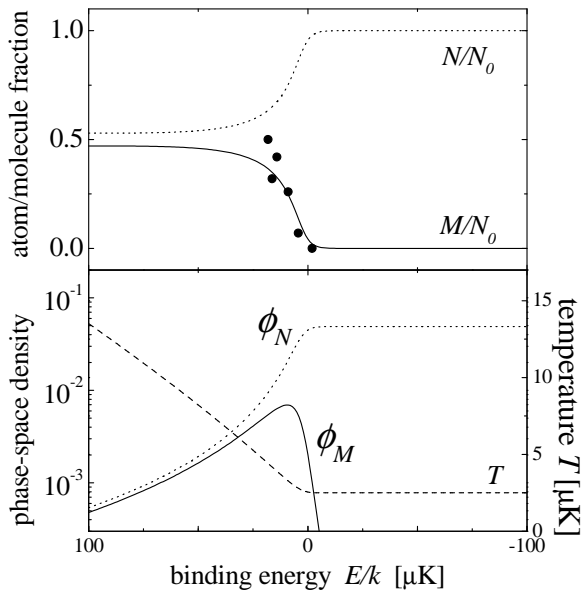}
\caption{Atom and molecule fraction in equilibrium starting with a
pure atom sample. Trap parameters are specified in the text. In
the upper figure, the atom fraction $N/N_0$ (dotted line) and the
molecule fraction $M/N_0$ (solid line) are calculated and compared
to the measurements (solid circles). The atom(molecule) fraction
of $50\%$ at large binding energy is accidental. In the lower
figure, the final phase-space density for molecules $\phi_M$
(solid line) and for atoms $\phi_N$ (dotted line) are shown
together with the temperature $T$ (dashed line). } \label{fig1}
\end{figure}

Based on the experiment parameters given in Ref.~\cite{joc03}:
axial (radial) trap frequency of $\omega_z$ $(\omega_r)=2\pi\times
260$ $(0.39)$ kHz, initial atom number $N_0=N'_0=1.25\times 10^6$,
molecule number $M_0=0$, and temperature $T_0=2.5\mu$K, we
calculate the molecule (atom) fraction, phase-space density and
temperature in thermal equilibrium, shown in Fig.~\ref{fig1}. A
maximum conversion efficiency of $2M/(N_0+N'_{0})=M/N_0\sim 50\%$
agrees with the experimental results ~\cite{joc04}.

Remarkably, the maximum conversion efficiency is reached at the
binding energy of $E > 20 \mu$K. This value greatly exceeds the
initial temperature of the sample and qualitatively explains the
displacement of the maximum recombination loss position by $\sim
150$G relative to the Feshbach resonance \cite{bou03, die02}.
However, a generalization of our treatment to quantum-degenerate
gas is necessary to predict the exact field shifts in those
experiments. For even more negative binding energy, an abrupt
turn-off of the molecule formation rate is observed and no
experimental data on the molecule number in thermal equilibrium is
available.

\begin{figure}
\includegraphics[width=3in]{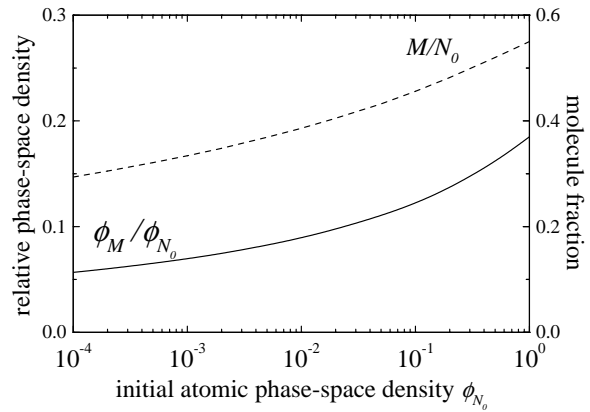}
\caption{Maximum conversion efficiencies in molecular phase-space
density $\phi_M$ and in molecule number $M$ from a pure atom
sample. The calculation is performed assuming an isotropic
harmonic trap, an initial atomic phase-space density $\phi_{N_0}$
and atom number $N_0=N'_0$. Notice that maximum molecular
phase-space densities $\phi_M$ (solid line) and numbers $M$
(dashed line) are obtained at different binding energies $E$.}
\label{fig2}
\end{figure}

To calculate the best atom-molecule conversion efficiency, we
assume initially $N_0=N'_0$ atoms are in an isotropic harmonic
trap with $Z_n=Z_{n'}=Z_m$, phase-space density
$\phi_{N_0}=\phi_{N'_0}$ and binding energy $E$, we calculate the
maximum molecular phase-space density $\phi_M$ or conversion
fraction $M/N_0$ by varying the binding energy. Over the range of
$\phi_{N_0}=10^{-4}\sim 1$, the maximum phase-space density of the
molecules $\phi_{M}$ is typically a factor of $5 \sim 20$ lower
than $\phi_{N_0}$ , while the maximum conversion fraction varies
from $30\%\sim55\%$, shown in Fig.~\ref{fig2}. This very weak
dependence on the initial atomic phase-space density is remarkable
and suggests that an efficient conversion of atoms into molecules
in thermal clouds is possible.

\section{Converting molecules to atoms}

\begin{figure}
\includegraphics[width=3in]{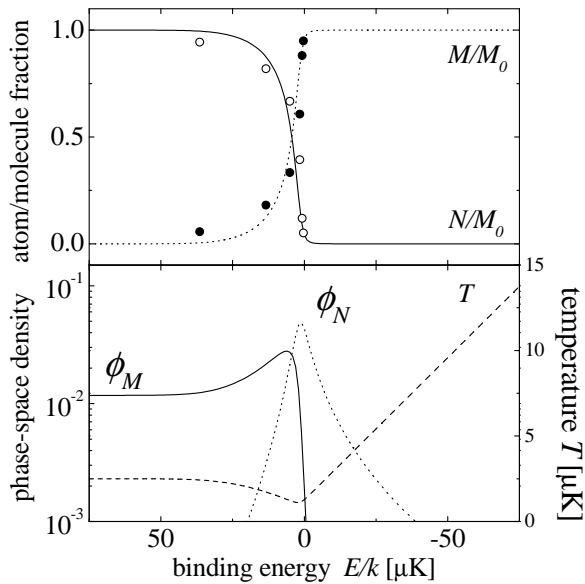}
\caption{Atomic and molecular fractions in equilibrium starting
with a pure sample of $M_0$ molecules. In the upper figure, the
fractions of molecules (solid line) and atoms (dotted line) are
calculated and compared to the experimental data: atoms (solid
circles) and molecules (open circles). In the lower figure, the
phase-space density of the molecules (solid line) and atoms
(dotted line) are shown together with the final temperature
(dashed line)} \label{fig3}
\end{figure}

Converting a pure molecule sample into atoms, as demonstrated in
Ref.~\cite{joc03}, shows different features. First, at large
negative binding energy, all molecules dissociate into atoms.
Beginning with $M_0=3\times 10^5$ molecules at $2.5 \mu$K with the
same trap parameters described earlier, we calculate the molecule
(atom) fraction $M/M_0$ ($N/M_0$) in thermal equilibrium, shown in
the upper figure of Fig.~\ref{fig3}. A full conversion from
molecules to atoms is achieved when the molecular state is high
above the scattering continuum. Shown together with the
calculation is the experimental data ~\cite{joc04}.

The agreement between the experiment and calculation is excellent
as there are no free parameters in the calculation. Both the
experiment and calculation support the possibility to dissociate
molecules into atoms even when the binding energy is positive and
large compared to the initial molecular temperature. This is due
to the much larger phase space of the atomic scattering continuum
as compared to that of the molecular bound state.

Furthermore, a gain in phase-space density after thermalization is
predicted  at small positive binding energies where temperature
drops and phase-space density peaks up, shown in the lower figure
of Fig.~\ref{fig3}. This cooling occurs due to two different
processes. First, the dissociation process is endoergic and
reduces the total external energy. Second, the total particle
number, $N+N'+M$, increases after molecules dissociate and further
reduces the mean energy per particle. When the equilibrium is
reached, we find a gain in molecular (atomic) phase-space density
of 2.4 (4.2), compared to that of the initial molecular
phase-space density. This cooling effect is studied in more detail
in Section V.

\section{Novel evaporative cooling in an atom-molecule mixture}

In conventional evaporative cooling schemes, energetic particles
are removed from the trap. The remaining particles rethermalize
and acquire a lower temperature and higher phase-space density.

\begin{figure}
\includegraphics[width=3in]{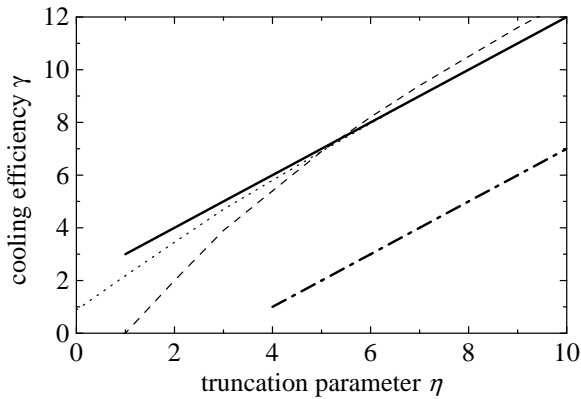}
\caption{Evaporation efficiency in an atom-molecule mixture.
Evaporation efficiency $\gamma$ during the thermalization process
is evaluated for $\eta=E/kT_0=10^{-4}$ (dashed line), and
$\eta=10^{0}$ (dotted line). Estimations based on the proposal
(solid line) and on conventional evaporation (dash-dotted line)
are also shown for comparison. No collision loss is considered in
the calculation.} \label{fig4}
\end{figure}

The molecule-atom conversion process resembles the above process.
By locating the atomic scattering continuum above the molecular
state with positive binding energy, molecules with higher thermal
energy are more probable to collisionally dissociate into atoms.
While these particles are immediately removed in the conventional
evaporative cooling recipe, the atoms from the dissociated
molecules have a greatly reduced thermal energy and a
rethermalization among these atoms and the remaining molecules is
advantageous to cool the sample. To evaluate the cooling
performance, we consider an initially pure sample of $M_0$
molecules in an isotropic harmonic trap with phase-space density
$\phi_{M_0}$, temperature $T_0$ and binding energy $E=\eta kT_0$,
where $\eta$ defines a truncation parameter, in analogy of the
energy cutoff in the conventional evaporative cooling scheme.
After the equilibrium is reached, we calculate the mean
evaporative cooling efficiency based on
$\gamma=-$ln$(\phi_{M}/\phi_{M_0})/$ln$(M/M_{0})$ \cite{ket96} for
a wide range of truncation parameter $\eta$, shown in
Fig.~\ref{fig4}, where $\phi_{M}$ $(M)$ is the molecular
phase-space density (number) in equilibrium. At constant $\eta$,
we find the atom-molecule thermalization permits a better cooling
efficiency $\gamma$ than does the conventional one.

On this basis, we propose a scheme to continuously cool molecules
toward high phase-space density and provide a quantitative
estimate on its performance. Given a harmonically trapped molecule
sample with negligible collisions loss, temperature $T$ and total
external energy $K$, we control the molecular binding energy
according to the cloud temperature and truncation parameter
according to $E=\eta kT$. We assume atoms resulting from molecule
dissociation are quickly thermalized with the sample before they
are removed. The removal of only atoms from an atom-molecule
mixture is demonstrated in Ref.~\cite{joc03}. During the
thermalization process, a small molecular fraction $\epsilon$ that
dissociates will lower the total external energy by $\delta
K=-K\epsilon\eta/3$; next, the thermalization process shares the
energy among the remaining $1-\epsilon$ molecules and $2\epsilon$
atoms and lowers the temperature by $\delta
T=-T\epsilon(\eta+3)/3$. Given the phase-space density in a
harmonic trap $\phi_M=MZ_m^{-1}\propto MT^{-3}$, we obtain the
evaporation efficiency as $\gamma=\eta+2$. The $\eta+2$ factor
comes from the increase in particle number, together with the
endoergic nature of the dissociation process. Compared to the
conventional evaporation with $\gamma_c=\eta+k-4$ \cite{ket96},
where $0<k<1$, the new scheme with, say, $\eta=4$ can have an
evaporative cooling performance comparable to that in conventional
method with $10>\eta>9$, shown in Fig.~\ref{fig4}.



However, this evaporation method involves additional atom-molecule
conversion and thermalization steps and can potentially be slow
due to either a slow thermalization rate or atom-molecule
conversion rate. In the two-component fermion system, in
particular, a large s-wave scattering length and elastic collision
cross section for atom-molecule  (molecule-molecule) scattering
near an atom-atom Feshbach resonance are expected \cite{pet03,
shly}. A fast thermalization rate on the order of $(1ms)^{-1}$ is
possible from recent experiments. The conversion rate from
molecules to atoms, however, depends on the details of the
atom-molecule interaction mechanism and will be discussed in Sec.
VI.

\section{Reaction mechanisms and evaporation rate}
To estimate the conversion rate or evaporation rate, specific
molecule formation and dissociation mechanism need to be
identified. We will consider Li system as a specific example.

In a dilute gas with negative binding energy, molecules dissociate
according to the following process,

\begin{eqnarray}
\mathrm{Li}_2 &\leftrightarrow & \mathrm{Li}+\mathrm{Li'}.
\label{eq3}
\end{eqnarray}

This spontaneous process is dominant since it is density
independent. The molecular dissociation rate in this case is
characterized by the Feshbach resonance linewidth \cite{paul}. In
typical experiments, typical Feshbach resonance linewidth of
$0.1$MHz $\sim 10$MHz indicates a very short dissociation time of
$\leq 1\mu$s.

When the binding energy is tuned to positive values $E>0$, the
process described by Eq.~(\ref{eq3}) is forbidden due to the
energy and momentum conservation. The leading binary processes to
dissociate molecules and the corresponding reversed ternary or
quaternary processes to form molecules are

\begin{eqnarray}
\mathrm{Li}_2 + \mathrm{Li} &\leftrightarrow & \mathrm{Li} + \mathrm{Li'} + \mathrm{Li} \label{eq4}, \\
\mathrm{Li}_2 + \mathrm{Li}_2 &\leftrightarrow & \mathrm{Li}_2 + \mathrm{Li} + \mathrm{Li'} \label{eq5} \\
\mathrm{Li}_2 + \mathrm{Li}_2 &\leftrightarrow & \mathrm{Li} +
\mathrm{Li'} + \mathrm{Li} + \mathrm{Li'}, \label{eq6}
\end{eqnarray}

\noindent Notice that the dissociation (formation) processes are
shown from left to right (right to left) and are allowed only when
the total energy of the incident channel is sufficient to support
the internal energy of the outgoing channel.

In thermal equilibrium, each molecule formation rate is exactly
balanced by the corresponding dissociation rate, due to detailed
balancing.  For the above processes, we have
$\overline{C_1mn}=\overline{R_1n^2n'}$,
$\overline{C_2m^2}=\overline{R_2mnn'}$ and
$\overline{C_3m^2}=\overline{R_3n^2n'^2}$, respectively, where
$C_i$ $(R_i)$ are the associated dissociation (formation) rate
coefficients, $n$, $n'$ $(m)$ are the atomic (molecular) densities
and $\bar{X}$ denotes the averaged value of $X$ in a canonical
ensemble. To directly relate the dissociation and formation
coefficients, we assume $n$ $(n')$ atoms and $m$ molecules are
uniformly distributed in a box with unity volume. The single
particle partition function is given by
$Z_n=Z_n'=2^{-3/2}Z_m=\lambda_{dB}^{-3}$, where
$\lambda_{dB}=h(2\pi m_0 kT)^{-1/2}$ is the thermal de Broglie
wavelength of the atom, $m_0$ atomic mass and $h$ Plank's
constant. Using Eq.~(\ref{eq1}), we obtain the relationship
between the coefficients in processes described by
Eq.~(\ref{eq4}), Eq.~(\ref{eq5}) and Eq.~(\ref{eq6}):

\begin{eqnarray}
\bar{C_1}&=&\bar{R_1}h^{-3}(\pi m_{0}kT)^{3/2}e^{-E/kT}, \label{eq7} \\
\bar{C_2}&=&\bar{R_2}h^{-3}(\pi m_{0}kT)^{3/2}e^{-E/kT}, \label{eq8}\\
\bar{C_3}&=&\bar{R_3}h^{-6}(\pi m_{0}kT)^3e^{-2E/kT}. \label{eq9}
\end{eqnarray}

\noindent The above ``formation-dissociation" relation reveals the
rate coefficient when the counterpart coefficient is calculated or
measured.

In samples which consist of mostly atoms, the dominant
atom-molecule conversion process is given in Eq.~(\ref{eq4}). The
formation reaction is widely studied in cold atoms systems and is
called three-body recombination \cite{fed96, esr01, pet03}. For a
two-component fermionic system, the rate coefficient is
$\bar{R_1}=167a^6kT/\hbar$ for $E/kT \gg 1$ \cite{pet03}. Using
the relationship $E=\hbar^2/m_{0}a^2$, and Eq.~(\ref{eq7}), we
immediately obtain the dissociation coefficient $\bar{C_1}=3.75
\hbar^2m_{0}^{-3/2}(kT)^{5/2}E^{-3}e^{-E/kT}$ for $E/kT \gg 1$
\cite{deconv}.

The situation is different when we start with a pure sample of
molecules, since the process in Eq.~(\ref{eq4}) does not happen in
the absence of atoms. Comparing Eq.~(\ref{eq8}) and
Eq.~(\ref{eq9}), we identify Eq.~(\ref{eq5}) as the dominant
dissociation process at low temperature since $\bar{C_3}$ is
exponentially suppressed relative to $\bar{C_2}$. This suppression
can be understood as the molecule-molecule collision energy in
Eq.~(\ref{eq6}) should be sufficiently high to support four atoms
in the continuum, while in Eq.~(\ref{eq5}), only two atoms are in
the continuum.

In contrast to that in Eq.~(\ref{eq4}), the formation process in
Eq.~(\ref{eq5}) involves scattering of three non-identical
particles: Li$_2$, Li and Li' and is expected to scale as
$R_2\sim\epsilon^0$ at low collision energy $\epsilon$ in the
three-body scattering channel. Consequently, dissociation
coefficient scales as $C_2\sim\epsilon^2$. Compared to $R_1\sim
\epsilon$ and $C_1\sim \epsilon^3$ \cite{deconv}, the dominant
conversion process in an atom-molecule mixture at low temperature
limit is actually Eq.~(\ref{eq5}), which involves four atoms. A
detailed calculation will be necessary to quantitatively determine
either $C_2$ or $R_2$.

Knowing the possible dissociation mechanism for molecules, we
estimate the speed of the proposed evaporative cooling based on
Eq.~(\ref{eq5}), in which the rate coefficients are known.
Given the evaporation parameter of $\eta=E/kT=4$, temperature of
$2.5\mu K$, and a small atom fraction with density $n=2\times
10^{11} cm^{-3}$, we obtain an evaporation rate of $\bar{C_1}n$=
(2.5s)$^{-1}$, which is indeed much slower than the two-body
collision rate. However, in conjunction with the predicted
evaporation efficiency of $\gamma \sim 5$, we expect an increase
of molecular phase-space density by 5 orders of magnitude can be
achieved in $6 s$, which is still within the lifetime of the
molecular cloud $10 s$ reported in Ref. \cite{joc03}. Furthermore,
during the evaporation process, temperature decreases and the
dissociation process in Eq.~(\ref{eq5}) will eventually dominate
and speed up the evaporation.

\section{Conclusion}
We have described the equilibrium condition and conversion
mechanisms in a thermal mixture of atoms and molecules under the
assumption of negligible collisional loss. Near a Feshbach
resonance, the molecular state extends the phase space of that of
two atoms. A full thermalization in this extended phase space
provides the equilibrium condition given in Eq.~(\ref{eq1}) and
permits a quantitative estimation on the thermodynamical
properties of an atom-molecule mixture.

In particular, we calculate the atom-molecule conversion
efficiencies under various conditions and the results agree with
the experimental data very well \cite{joc03, joc04}. We show that
by properly locating the binding energy of the molecules, the
conversion of a significant fraction of atoms to molecules is
possible even in a thermal gas with low phase-space density. Our
result suggests an alternative scheme to reach quantum degeneracy:
first, convert the fermionic atoms into bosonic molecules; second,
evaporatively cool them to a molecular condensate and finally,
convert them back to fermionic atoms. The last process would allow
a creation of atomic Cooper pairs \cite{hol01, oha02}.

To cool the molecules to a molecular condensate, we suggest a
novel cooling method based on the endothermic molecule-atom
conversion process at positive binding energy. We estimate a much
higher evaporative cooling efficiency can be obtained as compared
to conventional cooling in magnetic traps or dipole traps.

To estimate the time scale of thermalization and atom-molecule
conversion, we identify a new collision process given in
Eq.~(\ref{eq5}), which dominates the atom-molecule conversion at
low temperature. The evaporation rate of the proposed scheme is
estimated and an increase of molecular phase-space density by more
than five orders of magnitude within the typical lifetime observed
in the Li$_2$ thermal gas is expected \cite{joc03}. This result
highlights great prospect of attaining a molecular Bose-Einstein
condensate.

\section*{Acknowledgements}
We thank G. V. Shlyapnikov, P. Zoller, P. S. Julienne, A. Recati
and the members of the Innsbruck lithium team for stimulating
discussions. This work is supported by the Austrian Science Fund
(FWF) within SFB 15 (project part 15) and by the European Union in
the frame of the Cold Molecules TMR Network under contract No.
HPRN-CT-2002-00290.


\begin{references}
\bibitem{joc03} S. Jochim, M. Bartenstein, A. Altmeyer, G. Hendl, C. Chin, J. Hecker Denschlag, and R. Grimm, cond-mat/0308095.
\bibitem{don02} E.A. Donley, N.R. Claussen, S.T. Thompson, and C.E. Wieman, Nature {\bf 417}, 529 (2002).
\bibitem{her03} J. Herbig, T. Kraemer, M. Mark, T. Weber, C. Chin, H.-C. N\"{a}gerl, and R. Grimm,
Science 301, xxxx(2003), published online 21 August
(10.1126/science.1088876).
\bibitem{dur03} S. D\"{u}rr, T. Volz, A. Marte, and G. Rempe, cond-mat/0307440.
\bibitem{reg03} C.A. Regal, C. Ticknor, J.L. Bohn, and D.S. Jin, Nature {\bf 424}, 47 (2003).
\bibitem{cub03} J. Cubizolles, T. Bourdel, S.J.J.M.F. Kokkelmans, G.V. Shlyapnikov, and C. Salomon, cond-mat/0308018.
\bibitem{str03} K.E. Strecker, G.B. Partridge, and R.G. Hulet, cond-mat/0308318.
\bibitem{chi03} C. Chin, A.J. Kerman, V. Vuleti\'{c}, and S. Chu, Phys.\ Rev.\ Lett.\ {\bf 90}, 033201 (2003).
\bibitem{hol01} M. Holland, S.J.J.M.F. Kokkelmans, M. Chiofalo, and R. Walser, Phys.\ Rev.\ Lett.\ {\bf 87}, 120406 (2001).
\bibitem{oha02} Y. Ohashi, and A. Griffin, Phys.\ Rev.\ Lett.\ {\bf 89}, 130402 (2002)
\bibitem{fes62} H. Feshbach, Ann. Phys. (N.Y.) {\bf 5}, 357 (1958), {\bf 19}, 287
(1962).
\bibitem{ino98} S. Inouye, M. Andrew, J. Stenger, H.-J. Miesner, D. Stamper-Kurn, and W. Ketterle, Nature {\bf 392}, 151 (1998).
\bibitem{tim99} E. Timmermans, P. Tommasini, M. Hussein, and A.K. Kerman, Phys.\ Rep.\ {\bf 315} 199 (1999).
\bibitem{van99} F.A. van Abeelen, B.J. Verhaar, Phys.\ Rev.\ Lett.\ {\bf 83}, 1550 (1999).
\bibitem{mie00} F.H. Mies, E. Tiesinga, P.S. Julienne, Phys.\ Rev.\ A {\bf 61}, 022721 (2000).
\bibitem{tim01} E. Timmermans, K. Furuya, P.W. Milonni, and A.K. Kerman, Phys.\ Lett.\ A. {\bf 285} 228 (2001).
\bibitem{fed96} P.O. Fedichev, M.W. Reynolds, and G.V. Shlyapnikov, Phys\. Rev.\ Lett.\ {\bf 77}, 2921 (1996).
\bibitem{esr01} B.D. Esry, C.H. Greene, and H. Suno, Phys.\ Rev.\ A {\bf 65}, 010705(R) (2001).
\bibitem{pet03} D. Petrov, Phys.\ Rev.\ A {\bf 67}, 010703(R) (2003).
\bibitem{kok03} S.J.J.M.F. Kokkelmans, G.V. Shlyapnikov, and C. Salomon,
cond-mat/0308018.
\bibitem{hu03} H. Hu, F. Yuan, and S. Qin, cond-mat/0308474.
\bibitem{kerson} K. Huang, {\it Statistical Mechanics} (Wiley, New York, 1987).
\bibitem{har02} K.M. O'Hara, S.L. Hemmer, S.R. Granade, M.E. Gehm, J.E. Thomas, V. Venturi, E. Tiesinga, and C.J.
Williams, Phys.\ Rev.\ A {\bf 66}, 041401(R) (2002).
\bibitem{joc04} S. Jochim, Ph.D. thesis, Univ. Innsbruck, in preparation.
\bibitem{die02} K. Dieckmann, C.A. Stan, S. Gupta, Y. Hadzibabic, C.H. Schunck, and W. Ketterle, Phys.\ Rev.\ Lett.\ {\bf 89},
203201 (2002).
\bibitem{bou03} T. Bourdel, J. Cubizolles, L. Khaykovich, K.M.F. Magalh\~{a}es, S.J.J.M.F. Kokkelmans, G.V. Shlyapnikov, and C. Salomon, Phys.\ Rev.\ Lett.\ {\bf 91}, 020402 (2003).
\bibitem{ket96} W. Ketterle and N.J. Van Druten, Ad.\ At.\ Mol.\ Op.\ Phys.\ {\bf 37}, 181
(1996).
\bibitem{shly}  D.S. Petrov, C. Salomon, G.V. Shlyapnikov, cond-mat/0309010.
\bibitem{paul}  P.S. Julienne, private communication.
\bibitem{deconv} By deconvoluting the thermal averaging, we get $C_1=0.55
m^2a^7\hbar^{-5}\epsilon^3$, where $\epsilon$ is the mean kinetic
energy per atom in the outgoing channel. Our calculation confirms
the predicted dependence of $C_1$ on $\epsilon$ \cite{esr01}.

\end{references}
\end{document}